\documentclass[twoside,12pt]{article} 

\usepackage{a4wide}
\usepackage{isolatin1}
\usepackage{latexsym}

\newenvironment{myitemize}
{
  \begin{itemize}
    \setlength{\parsep}{\parskip}
    \addtolength{\itemsep}{-2.5pt}
}
{
  \end{itemize}
}
\newenvironment{mydescription}
{
  \begin{description}
    \setlength{\parsep}{\parskip}
    \addtolength{\itemsep}{-2.5pt}
}
{
  \end{description}
}
    
\newcommand{\SLAM}{SLAM}

\newcommand{\ToDo}[1]{
  \begin{quotation}
    \it
    \noindent
    \underline{\textbf{To do:}}
    #1
  \end{quotation}
}
\renewcommand{\ToDo}[1]{}

\newsavebox{\citaname}
\newenvironment{cita}[2]
{
  \sbox{\citaname}{\textsl{#2}}
  \hfill
  \begin{minipage}{#1}
    \begin{sl}}
{
  \\
    \mbox{}\hfill\usebox{\citaname}
    \end{sl}
  \end{minipage}}

\usepackage{amsfonts}

  \DeclareMathSymbol{\Bool} {\mathord}{AMSb}{"42}   
  \DeclareMathSymbol{\Nat}  {\mathord}{AMSb}{"4E}   
  \DeclareMathSymbol{\Int}  {\mathord}{AMSb}{"5A}   
  \DeclareMathSymbol{\Real} {\mathord}{AMSb}{"52}   
  \DeclareMathSymbol{\Rat}  {\mathord}{AMSb}{"51}   

\DeclareTextFontCommand{\SLAMntsFont}{\itshape\sffamily}
\def\makeNewSLAMnts#1#2{%
      \newcommand{#1}{\hbox{\SLAMntsFont{#2}\/}}}

\makeNewSLAMnts{\SLAMclassdecl}{class\_decl}
\makeNewSLAMnts{\SLAMclassexpr}{class\_expr}
\makeNewSLAMnts{\SLAMtypeorclassexpr}{type\_or\_class\_expr}
\makeNewSLAMnts{\SLAMvisibilitydegree}{visibility\_degree}
\makeNewSLAMnts{\SLAMinheritancedecl}{inheritance\_decl}
\makeNewSLAMnts{\SLAMcasedecl}{attribute\_case\_decl}
\makeNewSLAMnts{\SLAMargstypedecl}{args\_type\_decl}
\makeNewSLAMnts{\SLAMfunctiondecl}{function\_decl}
\makeNewSLAMnts{\SLAMconstructordecl}{constructor\_decl}
\makeNewSLAMnts{\SLAMobserverdecl}{observer\_decl}
\makeNewSLAMnts{\SLAMmodifierdecl}{modifier\_decl}
\makeNewSLAMnts{\SLAMexpression}{expression}
\makeNewSLAMnts{\SLAMarguments}{arguments}
\makeNewSLAMnts{\SLAMfunctiondef}{function\_def}
\makeNewSLAMnts{\SLAMidentifierlist}{identifier\_list}

\DeclareTextFontCommand{\SLAMtsFont}{\scshape}
\def\makeNewSLAMts#1#2{%
      \newcommand{#1}{\hbox{\SLAMtsFont{#2}\/}}}

\makeNewSLAMts{\SLAMIdentifier}{identifier}

\DeclareTextFontCommand{\SLAMKeywordFont}{\bfseries\sffamily}
\def\makeNewSLAMKeyword#1#2{%
      \newcommand{#1}{\hbox{\SLAMKeywordFont{#2}\/}}}

\makeNewSLAMKeyword{\SLAMPrivate}{private}
\makeNewSLAMKeyword{\SLAMProtected}{protected}
\makeNewSLAMKeyword{\SLAMPublic}{public}
\makeNewSLAMKeyword{\SLAMClass}{class}
\makeNewSLAMKeyword{\SLAMIsa}{isa}
\makeNewSLAMKeyword{\SLAMCase}{case}
\makeNewSLAMKeyword{\SLAMSequence}{Seq}
\makeNewSLAMKeyword{\SLAMBool}{Bool}
\makeNewSLAMKeyword{\SLAMNat}{Nat}
\makeNewSLAMKeyword{\SLAMInt}{Int}
\makeNewSLAMKeyword{\SLAMReal}{Real}
\makeNewSLAMKeyword{\SLAMOf}{of}
\makeNewSLAMKeyword{\SLAMIf}{if}
\makeNewSLAMKeyword{\SLAMFunction}{function}
\makeNewSLAMKeyword{\SLAMConstructor}{constructor}
\makeNewSLAMKeyword{\SLAMMember}{member}
\makeNewSLAMKeyword{\SLAMObserver}{observer}
\makeNewSLAMKeyword{\SLAMModifier}{modifier}
\makeNewSLAMKeyword{\SLAMPre}{pre}
\makeNewSLAMKeyword{\SLAMPost}{post}
\makeNewSLAMKeyword{\SLAMResult}{result}
\makeNewSLAMKeyword{\SLAMAnd}{and}
\makeNewSLAMKeyword{\SLAMOr}{or}
\makeNewSLAMKeyword{\SLAMNot}{not}
\makeNewSLAMKeyword{\SLAMImp}{implies}
\makeNewSLAMKeyword{\SLAMDimp}{equiv}
\makeNewSLAMKeyword{\SLAMTrue}{true}
\makeNewSLAMKeyword{\SLAMFalse}{false}
\makeNewSLAMKeyword{\SLAMAccept}{accept}
\makeNewSLAMKeyword{\SLAMQuantifier}{quantifier}
\makeNewSLAMKeyword{\SLAMCheck}{check}
\makeNewSLAMKeyword{\SLAMAndCheck}{and\_check}
\makeNewSLAMKeyword{\SLAMEitherCheck}{either\_check}
\newcommand{\result}{\SLAMResult}

\makeNewSLAMKeyword{\SLAMIs}{is}
\makeNewSLAMKeyword{\SLAMInherits}{inherits}
\makeNewSLAMKeyword{\SLAMAny}{any}
\makeNewSLAMKeyword{\SLAMEndClass}{endclass}
\makeNewSLAMKeyword{\SLAMIn}{in}
\makeNewSLAMKeyword{\SLAMMap}{map}
\makeNewSLAMKeyword{\SLAMSum}{sum}
\makeNewSLAMKeyword{\SLAMExtension}{\mbox{$+$}}

\newenvironment{slam}
{
  \vspace{\topsep}
  \hspace{-9pt}
  \sffamily
  \begin{tabular}[t]{l}
}
{
  \end{tabular}
  \vspace{\topsep}
}

\newcommand{\SLAMindent}{\mbox{\mbox{}\hspace{1em}\mbox{}}}
\newcommand{\SLAMexpr}[1]{{\sffamily #1}}
\newcommand{\SLAMmathexpr}[1]{\mathsf{#1}}

\newcommand{\SLAMFunSym}{$\mathsf{\rightarrow}$}
\newcommand{\SLAMOftypeSym}{\SLAMexpr{:}}
\newcommand{\SLAMOftype}{\SLAMexpr{$\mathsf{\in}$}}
\newcommand{\SLAMLabelSym}{\SLAMexpr{\&}}
\newcommand{\SLAMClassDecl}[1]{\SLAMexpr{\SLAMClass\ #1 \SLAMIs}}
\newcommand{\SLAMClassExpr}[2]{\SLAMexpr{#1\ #2}}
\newcommand{\SLAMInheritanceDecl}[2]{\SLAMexpr{#1\ \SLAMIsa\ #2}}
\newcommand{\SLAMCaseDecl}[3]{\SLAMexpr{#2\ #3\ \SLAMOftype\ #1}}
\newcommand{\SLAMLabelDecl}[3]{\SLAMexpr{#1\ \SLAMLabelSym\ #2\ \SLAMOftype\ #3}}
\newcommand{\SLAMFunctionDecl}[3]{\SLAMexpr{\SLAMFunction\ #1\ \SLAMOftypeSym\ #2\ \SLAMFunSym\ #3}}
\newcommand{\SLAMConstructorDecl}[3]{\SLAMexpr{\SLAMConstructor\ #2\ \SLAMOftypeSym\ #3}}
\newcommand{\SLAMObserverDecl}[4]{\SLAMexpr{\SLAMObserver\ #2\ \SLAMOftypeSym\ #3\ \SLAMFunSym\ #4}}
\newcommand{\SLAMModifierDecl}[3]{\SLAMexpr{\SLAMModifier\ #2\ \SLAMOftypeSym\ #3}}

\newcommand{\SLAMPreDef}[1]{\SLAMPre\ :- #1}
\newcommand{\SLAMFnDef}[3]{\SLAMexpr{#1\ #2\ = #3}}
\newcommand{\SLAMMemDef}[3]{\SLAMexpr{#1.#2\ = #3}}
\newcommand{\SLAMPostDef}[1]{\SLAMPost\ :- #1}

\newcommand{\SLAMFnCall}[2]{\SLAMexpr{#1\ #2}}
\newcommand{\SLAMMemFnCall}[3]{\SLAMexpr{#1.#2\ #3}}

\newcommand{\SLAMSol}[1]{\SLAMexpr{$\mathsf{\rhd}$ #1}}
\newcommand{\SLAMRecord}[1]{\SLAMexpr{\{#1\}}}

\newcommand{\SLAMSeqType}[1]{\SLAMexpr{\SLAMSequence\ #1}}
\newcommand{\SLAMSeq}[1]{\SLAMexpr{[#1]}}

\newcommand{\SLAMQFExpr}[5]{\SLAMexpr{#1\ #2\ \SLAMIn\ #3\ $|$ #4 $\mathsf{\bullet}$ #5}}
\newcommand{\SLAMQExpr}[4]{\SLAMexpr{#1\ #2\ \SLAMIn\ #3 $\mathsf{\bullet}$ #4}}

\newcommand{\SLAMTraversalDecl}[2]{\SLAMexpr{#1\ $\leadsto$ #2}}

\newcommand{\SLAMTraversal}[1]{\SLAMexpr{\SLAMSeq{#1}}}

\newcommand{\gargs}[1]{\overline{#1}}

\newcommand{\eclassgen}[1]{\vdots\\\SLAMEndClass}
\newcommand{\eclass}[1]{\SLAMEndClass}

\makeNewSLAMKeyword{\freerecord}{free\_record}

\newcommand{\aquan}{$\Xi$}
\newcommand{\ex}{$\exists$}
\makeNewSLAMKeyword{\existe}{exists}
\newcommand{\td}{$\forall$}
\makeNewSLAMKeyword{\ptodo}{forall}
\newcommand{\sm}{$\Sigma$}
\makeNewSLAMKeyword{\sumat}{sum}
\newcommand{\pd}{$\Pi$}
\makeNewSLAMKeyword{\product}{prod}
\newcommand{\sel}{$\exists^*$}
\makeNewSLAMKeyword{\select}{select}
\newcommand{\cnt}{${\cal N}$}
\makeNewSLAMKeyword{\recuento}{count}
\newcommand{\maxx}{${\cal M}$}
\makeNewSLAMKeyword{\maximo}{max}
\newcommand{\minn}{$\mu$}
\makeNewSLAMKeyword{\minimo}{min}
\newcommand{\mx}{${\cal M}^*$}
\makeNewSLAMKeyword{\maximizador}{maxim}
\newcommand{\mn}{$\mu^*$}
\makeNewSLAMKeyword{\minimizador}{minim}
\newcommand{\map}{{$\cal C$}}
\makeNewSLAMKeyword{\corresp}{map}
\newcommand{\fil}{{$\cal F$}}
\makeNewSLAMKeyword{\filtro}{filter}

\makeNewSLAMKeyword{\function}{function }
\makeNewSLAMKeyword{\private}{private}
\makeNewSLAMKeyword{\protected}{protected}

\makeNewSLAMKeyword{\member}{member}
\makeNewSLAMKeyword{\true}{true}
\makeNewSLAMKeyword{\false}{false}
\makeNewSLAMKeyword{\Sel}{sel }
\makeNewSLAMKeyword{\If}{if }

\newcommand{\qseqs}{{$\cal S$}}
\makeNewSLAMKeyword{\qseq}{seq}

\newcommand{\strav}[3]{\SLAMexpr{(#1, #2, #3)}}
\newcommand{\gentrav}{\SLAMexpr{tr$_1$, \ldots, tr$_n$}}

\makeNewSLAMKeyword{\acceptname}{accept}

\makeNewSLAMKeyword{\imitates}{imitates }
\makeNewSLAMKeyword{\by}{by}

\makeNewSLAMKeyword{\matchs}{matches }

\begin{document} 

\pagestyle{myheadings} 

\markboth{AADEBUG 2000}{Generation of and Debugging with Logical Pre and Postconditions}

\title{Generation of and Debugging with Logical Pre and
  Postconditions\footnote{In M. Ducass\'e (ed), proceedings of the
    Fourth International Workshop on Automated Debugging (AADEBUG
    2000), August 2000, Munich. COmputer Research Repository
    (http://www.acm.org/corr/), cs.SE/0101009; whole proceedings:
    cs.SE/0010035.}}

\author{
  Ángel Herranz Nieva \hspace{3em} Juan José Moreno Navarro\\
  Universidad Polit\'{e}cnica de Madrid \\
  Departamento LSIIS, Facultad de Inform\'{a}tica, \\
  Campus de Montegancedo, Boadilla del Monte, \\
  28660 Madrid, Spain, \\
  \texttt{mailto:\{aherranz,jjmoreno\}@fi.upm.es}}

\date{} 
\maketitle 

\begin{cita}{8cm}{C.A.R. Hoare}
  There are two ways of constructing a reliable system.  One is to
  make it so simple that there are no obvious deficiencies. The other
  one is to make it so complex that there are no obvious deficiencies.
\end{cita}

\begin{abstract} 
  
  \noindent
  
  This paper shows the debugging facilities provided by the \SLAM\ 
  system.  The \SLAM\ system includes i) a specification language that
  integrates \emph{algebraic specifications} and \emph{model-based
    specifications} using the object oriented model.  Class operations
  are defined by using rules each of them with logical pre and
  postconditions but with a functional flavour. ii) A development
  environment that, among other features, is able to generate readable
  code in a high level object oriented language.  iii) The generated
  code includes (part of) the pre and postconditions as assertions,
  that can be automatically checked in the debug mode execution of
  programs.
  
  We focus on this last aspect. The \SLAM\ language is expressive
  enough to describe many useful properties and these properties are
  translated into a Prolog program that is linked (via an adequate
  interface) with the user program. The debugging execution of the
  program interacts with the Prolog engine which is responsible for
  checking properties.
\end{abstract} 

\section{Introduction}

The paper presents the debugging subsystem of the ongoing project
\SLAM. The \SLAM\ system allows the user to specify a program in a
very high level specification language. It is an object oriented
formal specification language that integrates \emph{algebraic
  specifications} and \emph{model-based specifications}. An algebraic
specification, like those proposed by the OBJ language family
\cite{introOBJ,introFOOPS,maude} or Larch-LSL \cite{Guttag91a}, is
self-contained and independent of the chosen types in the code.
Basically a system is described by a decomposition in data
types/classes. The behaviour of an operation is described in terms of
other operations. On the other hand, specification languages based on
abstract models, as Z \cite{JBWordsworth:Z},
VDM \cite{CBJones:VDM} or Larch interfaces languages \cite{Guttag91a},
have a series of primitive mathematical domains (sets, sequences,
tables, etc) to model in an abstract way all data identified during
the process of development.  Specification of operations uses
(explicitly or implicitly) two predicates (pre/postconditions) that
describe the relationship between the input and the output by means of
a logic formula.

As we will see, \SLAM\ unifies both kind of languages by specifying
operations by logical pre and postconditions, but restricting logical
formulae to a computable view of quantifiers as traversal operations
on data.

The language is the heart of a system that integrates all the stages
in the programming process: analysis, design, implementation,
documentation and validation.

A \SLAM\ program is not directly stored in a text file. The
development environment allows to declare the full specification in a
convenient way. Different \SLAM\ components are different views of the
system and they can be edited independently. In order to facilitate
the understanding of \SLAM\ we will show \SLAM\ elements with a syntax
that does not necessarily correspond neither with an internal
representation nor the system presentation. They are generated via a
friendly interactive interface, so the reader should not pay too much
attention to the concrete syntax presented in this paper.

The \SLAM\ system is able to generate executable and readable code in
a high level object oriented language. Furthermore, the code contains
runtime checkable assertions (i.e. the pre and postconditions of
operations) that will help in the declarative debugging process of the
program.  The user can change the automatically generated code (for
instance, to make it more efficient) but the assertions can be reused
to ensure that the new code behaves correctly.

The assertions are complex logical formulae, so they are, in
principle, difficult to check. The \SLAM\ system generates a Prolog
program to check them.  The Prolog program is linked with the C++
program to achieve the desired effect.

A similar goal is present in some related works:
Anna \cite{Luckham90} is an Ada extension that includes facilities
for formally specifying the intended behaviour of Ada programs,
and the runtime checking of them. The assertions are limited to
Ada constructions and they work mainly with algebraic specifications.
In \cite{assrt-framework-jicslp98ws} a framework for assertion-based
debugging is presented: assertions are written using
Prolog and a type language so the user cannot specify properties in
such an abstract way as in \SLAM.  Other related works are
the {\em result checking} approach
\cite{result-checking} (the user must specify a new
algorithm in order to check the result of the original one),
and {\em relative debugging} \cite{Abramson:1996:RDN} 
(where it is possible to compare the execution of two programs 
with the same functionality)
Disadvantages appear clear, both algorithms/programs might be wrong and some
error might be undetectable.

>From the debugging point of view, the most novel feature of our proposal  
is that the specification language \SLAM\ is designed as a trade-off
between a high expressiveness and the possibility of compilation.
It serves both for specifying many useful properties about programs and, at the
same time, to generate executable and readable prototypes. It is 
possible because of the computable behaviour of quantifiers
thanks to the object oriented structure of the language.

We present here just a small subset of \SLAM\ features.  The
discussion of full \SLAM\ is beyond the aims of the paper and can be
found in the \SLAM\ webpage \cite{slam:web}. We focus on the most
significant \SLAM\ features and how to obtain the Prolog code to check
the assertions. The translation from \SLAM\ to Prolog is fully
formalized.

The paper organization is the following: we present the language in
section \ref{slam}, compilation schemes into C++ and Prolog in section
\ref{compiling} and the way in which Prolog assertion code is linked with
the C++ program are presented in section \ref{check}. Section
\ref{future} presents conclusions and future work.

\ToDo{ More related work }

\section{The Specification Language}
\label{slam}

\SLAM\ is an object oriented specification language. In this paper we
present a functional (i.e. stateless) version of the language. A
\SLAM\ program consists in a collection of user defined and predefined
classes. A \SLAM\ execution is an expression involving one or more
\SLAM\ objects and function (method) calls.  The next sections
describe the main \SLAM\ components.

\subsection{Classes}

In \SLAM\ a class is defined by specifying its properties. These
properties declare and define all the characteristics of the class.  A
class is merely declared by giving a class name.

A class can declare \emph{internal attributes}.  Attributes specify
the internal representation of instances of the class. In \SLAM\ 
attributes are not defined explicitly as in most of the imperative
object oriented languages. A more declarative construction
(\emph{algebraic types}) has been directly introduced into \SLAM\ to
indicate that a certain syntactical construction belongs to the class.

Every scheme of declaration of attributes defines an internal hidden
\emph{attribute constructor}.  The attribute constructor and its
arguments define an \emph{internal representation} of the object and
specification of operations can be written based on that model.  For
example, if we need a class to model 2D points, we can define the
following class:

\begin{slam}
\SLAMClassDecl{\SLAMClassExpr{Point}{}}\\
\SLAMindent\SLAMCaseDecl{Point}{Cartesian}{(Real,Real)}
\end{slam}

\noindent
and now expressions like \SLAMexpr{Cartesian (1,2.5)} are instances of
the class \SLAMexpr{Point}.

There is no restriction on the names of these attribute constructors,
it can even be omitted and replaced by a label\footnote{This label,
  expressed as \SLAMexpr{label \SLAMLabelSym}, can be omitted when
  there is just one alternative.}. For example, next examples show the
use of \emph{aggregation} in order to define the classes of segments
and stacks:

\begin{slam}
\SLAMClassDecl{Segment}\\
\SLAMindent\SLAMLabelDecl{seg}{\SLAMRecord{source : Point, destination : Point}}{Segment}
\end{slam}

\begin{slam}
\SLAMClassDecl{\SLAMClassExpr{Stack}{(Elem)}} \\
\SLAMindent\SLAMLabelDecl{contents}{\SLAMRecord{top : \SLAMNat, elements : \SLAMSeqType{(Elem)}}}{Stack}
\end{slam}

\ToDo{AHN: to explain the difference between labels and constructors.
  I don't know what it is.}

\ToDo{AHN: only one attribute set alternative means \emph{private
    inheritance}. The above definitions can be rewritten:

  \begin{slam}
    \SLAMClassDecl{Segment}\\
    \SLAMindent\SLAMPrivate\ \SLAMInheritanceDecl{Segment}{\SLAMRecord{source : Point, destination : Point}}
  \end{slam}

  \begin{slam}
    \SLAMClassDecl{\SLAMClassExpr{Stack}{(Elem)}} \\
    \SLAMindent\SLAMPrivate\ \SLAMInheritanceDecl{Stack}{\SLAMRecord{top : \SLAMNat, elements : \SLAMSeqType{(Elem)}}}
  \end{slam}
}

\noindent
The last example shows how classes can be parameterized. \SLAM\ 
provides bounded parametric polymorphism in the same style of Pizza
\cite{pizza}.  Some predefined \SLAM\ constructions have been used:

\begin{myitemize}

\item Record type construct with its standard semantics, for instance
  to define components indicating a source and a destination point.
  
\item Sequences, that are indexed collections of elements.  In the
  \SLAMexpr{Stack} class it is used to store the different elements in
  the stack.

\end{myitemize}

In contrast with other object oriented languages alternative
representations are allowed. In \SLAM\ a class can have several
alternative sets of attributes, the style is similar to how to
define new types in a functional language and quite related to
\emph{union types} or \emph{variant records} in some imperative
languages. A very similar construct can be found in Pizza
\cite{pizza}.

The point example can be extended to represent points in Polar
coordinates maintaining the Cartesian representation as another
alternative:

\begin{slam}
\SLAMClassDecl{Point}\\
\SLAMindent\SLAMCaseDecl{Point}{Cartesian}{(Real,Real)}\\
\SLAMindent\SLAMCaseDecl{Point}{Polar}{(Real,Real)}\\
\end{slam}

\noindent
Now, the expression \SLAMexpr{Polar (4.5,1.57)} also represents an
object that belong to the class \SLAMexpr{Point}.

The use of alternatives extends to the definition of recursive data
structures.  Let see a typical example of a recursive data structure
in \SLAM:

\begin{slam}
\SLAMClassDecl{\SLAMClassExpr{Tree}{(Elem)}}\\
\SLAMindent\SLAMCaseDecl{Tree}{Empty}{}\\
\SLAMindent\SLAMCaseDecl{Tree}{Node}{(Tree (Elem), Elem, Tree (Elem))}
\end{slam}

\ToDo{AHN: to explain how alternative representations are related to
  inheritance relationships.}

Up to now, there is no possibility to have references to object,
so it is not possible to include side effects. We plan to extend the
language in this vein.

\subsection{Class relationships}

\subsubsection{Inheritance}

\SLAM\ classes can be defined from the scratch by means of (multiple)
\emph{inheritance} from other class. In contrast with some other
object oriented languages attribute definitions can be refined by the
subclass by using the two usual kinds of inheritance:

\begin{myitemize}
  
\item Extension: New components are added to an attribute constructor.
\\ 
\begin{slam}
  \SLAMClassDecl{\SLAMClassExpr{ColouredPoint}{}}\\
  \SLAMindent\SLAMInheritanceDecl{ColouredPoint}{Point}\\
  \SLAMindent\SLAMCaseDecl{ColouredPoint}{Cartesian}{(Real,Real, Colour)}\\
\end{slam}
\mbox{}
\vspace{-10pt}
\mbox{}

\item Overriding (or refinement): The number of components/arguments
  in an attribute constructor remains the same but some of the
  components are replaced by subclasses of the original classes.
\\
\begin{slam}
\SLAMClassDecl{\SLAMClassExpr{ColouredSegment}{}}\\
\SLAMindent\SLAMInheritanceDecl{ColouredSegment}{Segment}\\
\SLAMindent\SLAMLabelDecl{seg}{\SLAMRecord{source : ColouredPoint, destination : ColouredPoint}}{ColouredSegment}
\end{slam}
\mbox{}
\vspace{-10pt}
\mbox{}

\end{myitemize}


\subsubsection{Aggregation}

Any class can be specified by declaring a representation based on
attributes that are instances of other classes. In the object oriented
model the structural relationship between classes is called
\emph{aggregation} or \emph{composition}.  

\subsubsection{Association}

\emph{Abstract classes} understood as \emph{Interfaces} are an
important component of \SLAM: they can be used to define an abstract
behaviour and other classes can indicate that they share this
behaviour. The main difference is that this ``indication'' can be done
by two alternative mechanisms. One is \emph{inheritance} and the user
need to supply the missing code or to reuse it.  The other one is {\em
  imitation}.  A class \SLAMexpr{The\_Class} can \emph{imitate} (implement in
Java terminology) the behaviour of another abstract class
\SLAMexpr{Abstract\_Class} by matching all its components (attributes
if any, functions) by components of \SLAMexpr{The\_Class}.

Abstract classes and imitation provides the full power of theories (in
OBJ terminology) as well as type classes (\`a la Haskell \cite{haskell:gentle, Haskell98}) playing a
more powerful role than C++ templates.

\subsubsection{Classes as types}

A class defines a datatype. In particular, it is clear that it
represents an algebraic
datatype when we move labels to constructors and different
alternatives are collected in an union type.  In the presence of
inheritance, type definitions are expanded.  Using Haskell
\cite{Haskell98} notation, the previous classes can be translated to
the following types, where constructors names are prefixed by the
class name to avoid name clashing:

\begin{small}
\begin{verbatim}
data Point1 = Point1Cartesian (Real, Real)
data Point2 = Point2Cartesian (Real, Real) | Point2Polar (Real, real)
data Segment = SegmentSeg (Point1, Point1)
data Stack a = StackContents (Int, Seq a)  % With an adequate definition of Seq
data Tree a = TreeEmpty | TreeNode (Tree a, a, Tree a)
data ColouredPoint = ColouredPointCartesian (Real, Real, Colour)
data ColouredSegment = ColouredSegmentSeg (ColouredPoint, ColouredPoint)
\end{verbatim}
\end{small}

\subsection{Predefined classes and relationships}

The \SLAM\ system includes a good number of predefined classes as well
as some predefined relations between them. Examples of predefined
classes are numbers, booleans, characters, dates, angles, strings,
records, sequences, lists, dictionaries, etc. An example of the
abstract class hierarchy comes from the \SLAMexpr{Collection} class.
The following figure shows the behavioural organization of these classes,
where the layout reflects inheritance:

\begin{sf}
\begin{small}
\begin{tabbing}
~\hspace{0.5cm}~ \= ~\hspace{0.5cm}~ \= ~\hspace{0.5cm}~ \= ~\hspace{0.5cm}~ \= ~\hspace{0.5cm} \kill
\>\>\>\>OrderedIdxCollection~\hspace{0.5cm}\={\rm\em As above, but respecting the order of elements} \kill
\>Collection \\
\>\>TravCollection \>\>\>{\rm\em Traversable collections} \\
\>\>FiniteTravCollection \>\>\>{\rm\em Finite traversable collections}\\
\>\>\>Set \\
\>\>\>Multiset \\
\>\>\>IndexedCollection \>\>{\rm\em Indexed collections. They are traversable}\\
\>\>\>\>OrderedIdxCollection \>{\rm\em As above, but respecting the order of elements}
\end{tabbing}
\end{small}
\end{sf}

\subsection{Traversals}

Any class that represents a collection of elements could specify a
\emph{traversal}. A traversal is either a \emph{simple enumeration} or
a sequence of traversals.

\begin{myitemize}
  
\item A \emph{simple enumeration} of a class is a triple
  \strav{first}{next}{inside}, where \SLAMexpr{first} is a class
  function that returns the initial element of the traversal,
  \SLAMexpr{next} is a class function that eliminates the first
  element and prepares it to give the following element of the
  traversal, and \SLAMexpr{inside} is a class boolean function
  indicating if the element already belongs to the traversal.

  \begin{myitemize}
    
  \item Arithmetic ranges \SLAMexpr{\SLAMSeq{n..m}} are objects in
    \SLAM, and can be seen as examples of simple traversals:
    \SLAMexpr{first} returns \SLAMexpr{n}, \SLAMexpr{next} consist in
    adding one to the lower limit \SLAMexpr{\SLAMSeq{n+1..m}}, and
    \SLAMexpr{inside} is the function that checks if the lower limit
    is lesser or equal than the upper one \SLAMexpr{n $\mathsf{\leq}$
      m}.
    
  \item A single element \SLAMexpr{x} is also a trivial example of a
    simple enumeration.
    
  \item A sequence is another example of single enumeration:
    \SLAMexpr{first} returns the first element of the sequence,
    \SLAMexpr{next} eliminates the head of the sequence and
    \SLAMexpr{inside} decides if there are still elements in the
    sequence.

\end{myitemize}

As soon as a class has a traversal defined (i.e. is an instance of the
\SLAMexpr{TravCollection} class) an object of the class can be used in
the place of a traversal.

\item A general traversal is a non empty sequence of traversal:
  \SLAMTraversal{\gentrav}. Traversals in the sequence can be of any
  class and the order is relevant, because the structure will be
  traversed using this order.  Typically, it contains simple
  traversals or objects of \SLAMexpr{TravCollection} classes. There is
  just one traversal per class, although several rules to define it
  can be used depending on the different shapes of the attributes.
  Traversals are specified in the following way:

\begin{slam}
\SLAMTraversalDecl{Shape ($\gargs{x}$)}{Traversal ($\gargs{x}$)}
\end{slam}

\noindent
where \SLAMexpr{Shape ($\gargs{x}$)} indicates the shape of the object,
and \SLAMexpr{Traversal} is a simple or general traversal.

An example of a traversal of a binary tree is the element at the root
(a simple enumeration), the left subtree (that can be traversed by the
same method) and the right subtree.

\begin{slam}
\SLAMTraversalDecl{Empty}{}\\
\SLAMTraversalDecl{Node (ls, root, rs)}{\SLAMTraversal{root, ls, rs}}
\end{slam}

Notice that changing the order of this definition means a different
way to traverse the tree (for instance, first traversing the left
subtree, then the root, and finally the right subtree by using the
traversal \SLAMTraversal{ls, root, rs}).
\end{myitemize}

A traversal can be interpreted as a way to generate a scheme of code
to traverse the data. A simple enumeration can be translated into a
single loop or a linear recursion, while a general traversal is moved
to a multiple recursive code (even with parallel execution).

\subsection{Class Operations}

\SLAM\ has a clear functional flavour, so class operations (methods)
can be represented by functions. For the shake of simplicity some
syntactical distinctions have been introduced to distinguish standard
object oriented aspects, although they can be inferred by the system.
The classification of operations is the following:

\begin{myitemize}
  
\item \emph{Object constructors.} An object constructor is a function
  designed to create new instances of a class. An object constructor
  declaration in class \SLAMexpr{Class} consists in the name of the
  constructor\footnote{There are no restrictions on constructor
    names.}, and the argument types:
\\
\begin{slam}
\SLAMConstructorDecl{Class}{f}{$\gargs{T}$}\\
\end{slam}
\hspace{1cm} means a function: \hspace{1cm} 
\begin{slam}
\SLAMFunctionDecl{f}{$\gargs{T}$}{Class}
\end{slam}
\mbox{}
\vspace{-30pt}
\mbox{}

\item \emph{Object observers.} Observers allow to access properties of
  an object without modifying it.  An observer of class
  \SLAMexpr{Class} of the form:
\\
\begin{slam}
\SLAMObserverDecl{Class}{f}{($\gargs{T}$)}{$R$}
\end{slam}
\hspace{0.6cm} means a function: \hspace{1cm} 
\begin{slam}
\SLAMFunctionDecl{f}{(Class,$\gargs{T}$)}{$R$}
\end{slam}
\mbox{}
\vspace{-20pt}
\mbox{}

\item \emph{Object modifiers.} Modifiers are designed to modify the
  value of an object. From the functional point of view a modifier of
  class \SLAMexpr{Class} of the form:
\\
\begin{slam}
\SLAMModifierDecl{Class}{f}{($\gargs{T}$)}
\end{slam}
\hspace{1cm} means a function:\hspace{1cm} 
\begin{slam}
\SLAMFunctionDecl{f}{(Class,$\gargs{T}$)}{Class}
\end{slam}
\mbox{}
\vspace{-20pt}
\mbox{}

\item \emph{Friend functions.} They are functions that involve two or
  more objects of the class and they have not special decoration to be
  declared.

\end{myitemize}

Function invocation for constructors, observers and modifiers is done
via the usual dot notation, i.e. function \SLAMexpr{f} of class
\SLAMexpr{Class} is called in the object \SLAMexpr{obj} by the
expression \SLAMexpr{obj.f ($\gargs{a}$)} or, in case of ambiguity, by
\SLAMexpr{obj.Class:f ($\gargs{a}$)}.

In \SLAM\ a function is specified by a set of \emph{rules}. Every rule
involves a \emph{precondition} that indicates if the rule can be
apply, a \emph{function call scheme}, and a \emph{postcondition} that
relates input and output.

The general form of a function specification is the following:

\begin{slam}
  \SLAMPreDef{$P (\gargs{x})$}\\
  \SLAMFnCall{f}{($\gargs{a})$}\\
  \SLAMPostDef{$Q (\gargs{x},\SLAMResult)$}
\end{slam}

\noindent
where $P (\gargs{x})$ is a \SLAM\ formulae involving variables of the
arguments ($\gargs{a}$) and $Q (\gargs{x},\SLAMResult)$ is another
\SLAM\ formula. The variable (and reserved word) \SLAMResult\ is
always used to represent the computed value of the function. The
formal meaning of that specification is the assertion of the following
fact:

\begin{displaymath}
P (\gargs{x}) ~\Rightarrow~ Q(\gargs{x},\SLAMmathexpr{f} (\gargs{a}))
\end{displaymath}

\noindent
Although \SLAM\ formulas and expressions will be defined in detail
later the reader can accept some examples on class \SLAMexpr{Point}:

\begin{slam}
  \SLAMConstructorDecl{Point}{PointAt}{(Real,Real)}\\
  \SLAMPreDef{\SLAMTrue}\\
  \SLAMFnCall{PointAt}{(x, y)}\\
  \SLAMPostDef{\SLAMResult\ = Cartesian (x, y)}
\end{slam}
\hspace{5em}
\begin{slam}
  \SLAMObserverDecl{Point}{CoordX}{()}{Real}\\
  \SLAMPreDef{\SLAMTrue}\\
  \SLAMMemFnCall{Cartesian (x,y)}{CoordX}{()}\\
  \SLAMPostDef{\SLAMResult\ = x}
\end{slam}

\begin{slam}
\SLAMFunctionDecl{Distance}{(Point,Point)}{Real} \\
  \SLAMPreDef{\SLAMTrue}\\
  \SLAMFnCall{Distance}{(Cartesian (x1,y1), Cartesian (x1,y2))} \\
 \SLAMPostDef{\SLAMResult\ = sqrt (sqr (x1 - x2) - sqr (y1-y2))} \\
\end{slam}

\noindent
As in VDM explicit function definitions are allowed. Even more,
unconditionally true preconditions can be skipped what provides more
concise and readable specifications.

\begin{slam}
  \SLAMObserverDecl{Point}{CoordY}{()}{Real}\\
  \SLAMMemDef{Cartesian (x,y)}{CoordY ()}{y}
\end{slam}


\begin{slam}
  \SLAMModifierDecl{Point}{MoveLeft}{(Real)}\\
  \SLAMMemFnCall{p}{MoveLeft}{(d)}\\
  \SLAMPostDef{(Distance (\SLAMResult, p) = d) \SLAMAnd\ 
    (p.CoordY() = \SLAMResult.CoordY())}
\end{slam} 

\noindent
A function specification only indicates the relationship between the
result and the arguments, but not necessarily a method to compute the
function. Such a method is called a \emph{solution} in \SLAM\ 
terminology. A solution always gives a \emph{computable expression} to
the function (i.e. to the \result\ variable in the postcondition).
Next section will describe exactly what is a computable \SLAM\ 
expression.

In those cases where the specification is not a solution, it can be
specified in the following way:

\begin{slam}
  \SLAMFunctionDecl{f}{($\gargs{T}$)}{Class}\\
  \SLAMPreDef{$P (\gargs{x})$}\\
  \SLAMFnCall{f}{($\gargs{a}$)}\\
  \SLAMPostDef{$Q (\gargs{x},\SLAMResult)$}\\
  \SLAMSol{Solution $(\gargs{x}$)}
\end{slam}

\noindent
where \SLAMexpr{Solution} is a computable \SLAM\ expression. For
instance, we could have:

\begin{slam}
  \SLAMModifierDecl{Point}{MoveLeft}{(Real)}\\
  \SLAMMemFnCall{p}{MoveLeft}{(d)}\\
  \SLAMPostDef{(Distance (\SLAMResult, p) = d) \SLAMAnd\ 
    (p.CoordY() = \SLAMResult.CoordY())} \\
  \SLAMSol{Cartesian (-p.CoordX (), p.CoordY ())}
\end{slam}

\subsection{SLAM formulae and quantifiers}

\SLAM\ formulas are basically logic formulas using the usual logical
connectives (\SLAMAnd\ - conjunction, \SLAMOr\ - disjunction,
\SLAMNot\ - negation, \SLAMImp\ - implication, and \SLAMDimp\ -
equivalence), predefined and user defined functions and predicates,
and \emph{quantifiers}.

\SLAM\ formulas are typed in a similar way than other expressions. In
fact, expressions and formulas share the syntax -- every formula is an
expression of type boolean.

\SLAM\ expressions can combine objects with its own operations.
Operations can be combined in any consistent way to produce new
expressions. Expressions can use quantifiers of adequate types.
Quantifiers are a key feature in \SLAM\ and extend the notion of
quantifier in logic. Quantifiers can compute not only the truth of an
assertion but also any other value. Moreover quantifiers can be
applied to any \emph{collection} object.  Examples of collections are
sets, sequences, multisets, lists, trees, etc. The basic reading of a
quantifier is the traversal of a collection object in order to apply
an operation to all the elements in the collection. Therefore, the
components of a quantifier are:

\begin{myitemize}
  
\item The quantifier itself that generalizes a basic binary operation.
  
\item The description of the collection which is traversed.
  
\item A filter (boolean expression) that indicates which elements in
  the collection are involved in the quantification. Filters are
  introduced to avoid the generation of many intermediate objects.
  
\item The quantified expression

\end{myitemize}

A quantifier is written in the following way:

\begin{center}
\SLAMQFExpr{\aquan}{x}{D}{\mbox{\textit{filter} (x)}}{\mbox{\textit{expression} (x)}}
\end{center}

\noindent
where \aquan\ is the quantifier symbol (that indicates the meaning of
the quantifier), \SLAMexpr{D} is a collection where the quantifier
ranges, \SLAMexpr{x} is the quantified variable,
\SLAMexpr{\textit{filter}} is a boolean expression that restricts the
elements of \SLAMexpr{D} that are taking into account for the
quantification and \SLAMexpr{\textit{expression}} is a function to
interact with the quantification.

We describe some predefined quantifiers, one for each kind,
although \SLAM\ includes all used in logic and many more: 

\begin{mydescription}
  
\item[Existential quantification:] (\existe\ or \ex): Decides if there
  is an element in \SLAMexpr{D} where \SLAMexpr{\textit{filter}} is true,
  verifying \SLAMexpr{\textit{expression}}. It is a generalization of the
  logical disjunction.
  
 \item[Universal quantification:] (\ptodo\ or \td): Decides if all the
   elements in \SLAMexpr{C} where \SLAMexpr{\textit{filter}} is true,
   verify \SLAMexpr{\textit{expression}}. It generalizes the logical
   conjunction.
  
\item[Summatory:] (\sumat\ or \sm): Sums all the values \SLAMexpr{\it
    expression (x)} where \SLAMexpr{x} ranges in the elements of
  \SLAMexpr{D} where \SLAMexpr{\textit{filter}} is true.  Obviously, the
  addition is generalized by this quantifier.
  
 \item[Productory:] (\product\ or \pd): Multiplies all the values
   \SLAMexpr{\textit{expression} (x)} where \SLAMexpr{x} ranges in the
   elements of \SLAMexpr{D} where \SLAMexpr{\textit{filter}} is true (by
   generalizing multiplication).
  
\item[Counting:] (\recuento\ or \cnt): Counts the elements in
  \SLAMexpr{D} where \SLAMexpr{\textit{filter}} is true, verifying
  \SLAMexpr{\textit{expression}}.
  
\item[Selection:] (\select\ of \sel): Returns any element in
  \SLAMexpr{D} where \SLAMexpr{\textit{filter}} is true, verifying
  \SLAMexpr{\textit{expression}}.
  
\item[Maximum:] (\maximo\ or \maxx): Returns the maximum of all the
  values \SLAMexpr{\textit{expression} (x)} where \SLAMexpr{x} ranges in
  the elements of \SLAMexpr{D} where \SLAMexpr{\textit{filter}} is true.
  
\item[Maximizer:] (\maximizador\ or \mx): Returns the value
  \SLAMexpr{y} such that \SLAMexpr{\textit{expression} (y)} is the maximum
  of all the values \SLAMexpr{\textit{expression} (x)} where \SLAMexpr{x}
  ranges in the elements of \SLAMexpr{D} where \SLAMexpr{\textit{filter}}
  is true.
  
 \item[Minimum:] (\minimo\ or \minn): Returns the minimum of all the
   values \SLAMexpr{\textit{expression} (x)} where \SLAMexpr{x} ranges in
   the elements of \SLAMexpr{D} where \SLAMexpr{\textit{filter}} is true.
  
 \item[Minimizer:] (\minimizador\ or \mn): Returns the value
   \SLAMexpr{y} such that \SLAMexpr{\textit{expression} (y)} is the minimum
   of all the values \SLAMexpr{\textit{expression} (x)} where \SLAMexpr{x}
   ranges in the elements of \SLAMexpr{D} where \SLAMexpr{\textit{filter}}
   is true.
  
 \item[Filter:] (\filtro\ or \fil): Returns an element of the same
   class of \SLAMexpr{D} but eliminating the elements that verify
   \SLAMexpr{\textit{expression}}.
  
 \item[Map:] (\corresp\ or \map): Returns an structure with the elements
   \SLAMexpr{\textit{expression} (x)} such that \SLAMexpr{x} ranges in the
   elements of \SLAMexpr{D} where \SLAMexpr{\textit{filter}} is true.

\item[Sequence constructor:] (\qseq\ or \qseqs): Constructs a sequence
with indexes in $D$ and elements the values of \SLAMexpr{\textit{expression} (x)}.

\end{mydescription}

\subsubsection{Computable \SLAM\ expression}

Any \SLAM\ expression can be computed provided that quantifiers are
applied to a \SLAMexpr{FiniteTravCollection} class. A
\SLAMexpr{FiniteTravCollection} is a collection class whose instances
can be traversed and they have a finite number of elements.  

\subsection{Asserting properties}

The \SLAM\ compiler is designed to produce executable and readable
imperative code. One of the main features of the resulting code is
that it contains
debugging annotations, similar to the \texttt{assert} directive of C.
The main idea is to include assertions to check pre and
postconditions. However, this is not always possible.  \SLAM\ can
generate checking code for formulas involving quantifications over
objects that can be finitely traversed, although it is not necessary -
for postconditions - that they assign a value to \result.

When a formula cannot be checked completely, the user (maybe later
with the help of the \SLAM\ compiler) can indicate three possible
situations.  For simplicity we apply them to the postcondicion but the
same annotations can be done for the precondition.

\begin{myitemize}
  
\item The full formula can be checked. This is done by adding the
  keyword \SLAMCheck\ to the formula. This is the case when the
  specification is yet a solution and in this case no annotation is
  needed. 
\\
\begin{slam}
  \SLAMFunctionDecl{f}{($\gargs{T}$)}{Class}\\
  \SLAMPreDef{$P (\gargs{x})$}\\
  \SLAMFnCall{f}{($\gargs{a}$)}\\
  \SLAMPostDef{\SLAMCheck\ $Q (\gargs{x},\SLAMResult)$}\\
\end{slam}
\mbox{}
\vspace{-10pt}
\mbox{}

\item Just a part of the formula can be checked. This means that it is
  a part of a conjunction, so we assume that first part of the formula 
  cannot be checked, and the second one is marked by the keyword
  \SLAMAndCheck.
\\
\begin{slam}
\SLAMFunctionDecl{f}{($\gargs{T}$)}{Class}\\
  \SLAMPreDef{$P (\gargs{x})$}\\
  \SLAMFnCall{f}{($\gargs{a}$)}\\
  \SLAMPostDef{non-checkable-property($\gargs{x}$, \SLAMResult) \SLAMAndCheck\ 
                    checkable-property($\gargs{x}$, \SLAMResult})
\end{slam}
\mbox{}
\vspace{-10pt}
\mbox{}

\item The formula cannot be checked at all or the part that can be
  checked is not significant. In this case, the user can provide an
  approximation of the property that is checkable by placing it after
  the keyword \SLAMEitherCheck.
\\
\begin{slam}
\SLAMFunctionDecl{f}{($\gargs{T}$)}{Class}\\
  \SLAMPreDef{$P (\gargs{x})$}\\
  \SLAMFnCall{f}{($\gargs{a}$)}\\
  \SLAMPostDef{non-checkable-postcondition($\gargs{x}$, \SLAMResult) 
              \SLAMEitherCheck\ checkable-approximation($\gargs{x}$, \SLAMResult)}
\end{slam}
\mbox{}
\vspace{-20pt}
\mbox{}

\end{myitemize}

\subsection{An additional example}

Let us show a more complete but simple example of a \SLAM\
specification.  The problem is to handle transactions between saving
banks. The input is a collection of transactions (with the source
bank, the destination bank and the amount) and the collection of bank
names. The function to specify must compute the final profit/debit for
each bank.

\begin{slam}
  \SLAMClassDecl{Transaction}\\
  \SLAMindent\SLAMPublic\ \SLAMLabelDecl{tran}{\SLAMRecord{source : String, destination : String, amount : Real}}{Transaction}
\end{slam}

\begin{slam}
  \SLAMClassDecl{Bank}\\
  \SLAMindent\SLAMPublic\ \SLAMLabelDecl{bank}{\SLAMRecord{name : String, amount : Real}}{Bank} 
\\
%
  \SLAMConstructorDecl{Bank}{MakeBank}{(String,Real)}\\
  \SLAMFnDef{MakeBank}{(n, a)}{\SLAMRecord{n, a}}
\end{slam}

\begin{slam}
  \SLAMClassDecl{CTransaction}\\
  \SLAMindent\SLAMLabelDecl{ctran}{\SLAMSeqType{(Transaction)}}{CTransaction}
\end{slam}

\begin{slam}
\SLAMObserverDecl{CTransaction}{FinalAmount}{\SLAMSeqType{(String)}}{\SLAMSeqType{(Bank)}}\\
\SLAMPreDef{banks $\neq$ \SLAMSeq{}}\\
\SLAMFnCall{FinalAmount}{(ctrans, banks)}\\
\SLAMPostDef{\SLAMCheck\
  \begin{slam}
    length (\SLAMResult) = length (banks) \SLAMAnd\\
    \SLAMQExpr{\ptodo}{i}{\SLAMSeq{1..length(\SLAMResult)}}{~}\\
    \SLAMindent\SLAMResult(i).name = banks(i) \SLAMAnd\\
    \SLAMindent\SLAMResult(i).amount =
    \begin{slam}
      (\SLAMQFExpr{\sumat}{t}{ctrans}{t.source = banks(i)}{t.amount}) -\\
      (\SLAMQFExpr{\sumat}{t}{ctrans}{t.destination = banks(i)}{t.amount})
    \end{slam}
    \vspace{-2.5ex}
  \end{slam}}
\\
\SLAMSol{
  \SLAMQExpr{\corresp}{n}{banks}{MakeBank (n,
    \begin{slam}
      (\SLAMQFExpr{\sumat}{t}{ctrans}{t.source = n}{t.amount}) -\\
      (\SLAMQFExpr{\sumat}{t}{ctrans}{t.destination = n}{t.amount}))
    \end{slam}}}
\end{slam}

\ToDo{To explain how can CTransaction access to internal atributes of
  Transaction and Bank}

\subsection{Additional features}

\SLAM\ includes some other features that are quite convenient for
program specification. However, due to the lack of space we have
omitted those characteristics that are less interesting for the
purpose of this paper. Among them we can mention: \emph{Selectors},
that extend patterns for hidden constructors of classes,
\emph{Internal laws}, to simplify the data stored in a class after any
operation, a broad notion of visibility in case of aggregation, etc.

\section{Compiling \SLAM\ programs}
\label{compiling}

\subsection{Compiling \SLAM\ to an imperative object oriented language}

One of the main features of the \SLAM\ is that it is able to generate
code from a specification.
We will use C++ (\cite{c++}) as the target language. However the ideas
are applicable to any language with similar characteristics (Java,
Smalltalk, etc.).  In this section we only sketch the main ideas for
the compilation because this is not the main goal of the paper. The
basis for the compilation are:

\begin{myitemize}
  
\item \SLAM\ classes are translated to C++ classes.
  
  An \SLAM\ class defines a declarative type via its attributes.  This
  type is translated into an imperative type. The C++ class contains a
  single attribute of this type. The construction of an imperative
  type from a declarative one is a key feature for the compilation.
  Constructors are translated to records, alternative definitions are
  moved to union types while recursive definitions need the inclusion
  of a pointer.
  
\item \SLAM\ class relationships are almost mimic by C++ class
  relationships.

  \begin{myitemize}

  \item Aggregation is already included in object oriented languages.
    
  \item Most of the power of inheritance on \SLAM\ is included in C++,
    except the inheritance modification of the state. Attribute
    constructor extension (or record extension) can be modelled by C++
    attribute extension (i.e. additional attributes can be supplied
    in the subclass), while state overriding can
    be implemented by attribute redefinition\footnote{In C++, when a
      subclass declares an attribute with the same name of an
      attribute of the superclass, this last one is hidden by the
      newest.}.
    
  \item Association can be simulated by inheritance and/or the use of
    templates.

  \end{myitemize}
  
\item \SLAM\ predefined classes are implemented by specifically
  designed C++ classes, although some of them need a
  special treatment by the compiler.
  
\item \SLAM\ functions are implemented by C++ methods in traversable
  classes. The \SLAM\ classification of operations are easy to map
  into C++ classification.
  
\item Traversals are implemented by including the operations
  \texttt{first}, \texttt{next}, and \texttt{inside} as class methods.
  Quantifiers are implemented by adequate loops.
  
\item An additional operation \texttt{serialize} is included in all
  the classes to store an element into a file using the XML format. 
  This operation is also automatically generated.

\end{myitemize}

\noindent
In our previous example, we can obtain automatically a C++ code
similar to the following one\footnote{Some parts have been skipped, and some
  simplifications like using arrays as the \SLAMSequence\ 
  implementation have been included to improve readability.}:

\begin{small}
\noindent
\begin{minipage}[t]{.4\textwidth}
\begin{verbatim}
// in file transaction.h 
class Transaction { 
 public: 
   String source; 
   String destination; 
   float amount; 

   void serialize ();
}; 
\end{verbatim}
\begin{verbatim}
// in file bank.h 
class Bank { 
  public: 
    String name; 
    float  amount; 

    void Bank (String n, float a) {
      name = n;
      amount = a;
    }
    void serialize ();
}; 
\end{verbatim}
\begin{verbatim}
// in file ctransaction.h 
class CTransaction { 
  public: 
    void FinalAmount (
      String banks [MaxBanks], 
      Bank result [MaxBanks]);
    void serialize ();

    Transaction trans [MaxTrans]; 
}; 
\end{verbatim}
\end{minipage}
\begin{minipage}[t]{.4\textwidth}
\begin{verbatim}
// in file ctransaction.cpp
void CTransaction::serialize () {
...
}

void CTransaction::FinalAmount (
  String banks [MaxBanks], 
  Bank result [MaxBanks])
{ 
  pre-check (FinalAmount (banks, MaxBanks)); 
  for (int i = 0; i <= MaxBanks; i++){ 
    float amount = 0; 
    for (int j = 0; j <= MaxTrans; j++){ 
      if (trans[j].source == banks [i]) 
        amount =+ trans[i].amount; 
      else if (trans[j].destination == banks[i]) 
        amount =- trans[i].amount; 
    } 
    result[i] = new Bank (banks [i], amount); 
  } 
  post-check (FinalAmount 
                (banks, MaxBanks, result))
} 
\end{verbatim}
\end{minipage}
\end{small}

This automatically generated code is not necessarily intended to be an
efficient code. In fact, it could be considered a prototype code,
although in many cases the code is good enough. In those situations,
or even in situations in which \SLAM\ cannot generate code, the user
can modify it and write an optimizing code.

In our previous example, the programmer can decide that it is better
to traverse first the transaction collection and then search the name
of the banks to update the result.  The new code is:

\begin{small}
\begin{verbatim}
void CTransaction::FinalAmount (String banks [MaxBanks], 
                                Bank result [MaxBanks]) { 
  pre-check (FinalAmount (banks, MaxBanks)); 
  for (int i = 0; i <= MaxBanks; i++){ 
    result[i].name = banks [i]; 
    result[i].amount = 0; 
  }
  for (int j = 0; j <= MaxTrans; j++){ 
    for (i = 0; i < MaxBanks; i++) {
      if (trans[j].source == result [i].name)
        result[i].amount =+ trans[i].amount; 
    }
    for (i = 0; i < MaxBanks; i++) {
      if (trans[j].destination == result [i].name) 
        result[i].amount =+ trans[i].amount;  // There is a bug here
    }
  } 
  post-check (FinalAmount (banks, MaxBanks, result))
} 
\end{verbatim}
\end{small}

Notice that the code contains a bug when the destination bank is
found.  Instead of decreasing the amount it is increased.  Program
assertions remain valid, so the bug will be detected when the program
is executed in the debug mode.

\ToDo{To explain the translation from sequences into arrays ``by the
  face''.}

\ToDo{To translate with exceptions!!! pre-check and post-check throws
  exceptions!!!}

\ToDo{\texttt{pre-check} doesn't compile.}

\subsection{Compiling \SLAM\ to Prolog}
\newcommand{\ciao}{Ciao/Prolog}

More relevant for the purpose of the paper is the compilation of
\SLAM\ programs and assertions into a Prolog piece of code.

We use the \ciao\ system \cite{ciao-novascience} as the target
language.  The main reasons are our knowledge of the system and our
easy access to the \ciao\ system developers, the availability of some
higher order features that simplify the implementation of
quantifiers, and the existence of an interface to connect C/C++ and
Prolog programs.  Anyway, the most important ideas presented here are
independent of the \ciao\ system and are applicable to any other
Prolog system.

We have already explained that the semantics of classes are algebraic
types.  Instances of classes will be translated using Prolog functors
as data constructors. This is the usual way in which other declarative
languages translates algebraic types into Prolog. The name of the
class is added to the name of the constructor in order to avoid
clashing in the presence of inheritance.

  

\newcommand{\solname}[2]{\texttt{'sol-}#2\texttt{'}}
\newcommand{\prename}[2]{\texttt{'pre-}#2\texttt{'}}
\newcommand{\postname}[2]{\texttt{'post-}#2\texttt{'}}
\newcommand{\readname}[1]{\texttt{'read-}#1\texttt{'}}
\newcommand{\quant}[1]{\texttt{'quan-}#1\texttt{'}}
\newcommand{\inname}[1]{\texttt{in}}
\newcommand{\casting}[1]{\texttt{'to-}#1\texttt{'}}
\newcommand{\translate}{\textit{translate}}
\newcommand{\transrule}{\textit{trans\_rule}}
\newcommand{\transexpr}{\textit{trans\_expr}}
\newcommand{\transterm}{\textit{trans\_term}}
\newcommand{\transpre}{\textit{trans\_pre}}
\newcommand{\transpost}{\textit{trans\_post}}
\newcommand{\transsol}{\textit{trans\_sol}}
\newcommand{\transinh}{\textit{trans\_inh}}
\newcommand{\transpred}{\textit{trans\_predefined}}
\newcommand{\transread}{\textit{trans\_read}}
\newcommand{\transtrav}{\textit{trans\_trav}}
\newcommand{\miss}{\textit{miss}}
\newcommand{\defines}{\textit{defines}}
\newcommand{\labs}{$\mathsf{\backslash}$}

A formal specification of the translation of function
definitions is given in figure \ref{trans}. Every function is compiled
into three predicates that implements different functionalities. For
every function $f$ with arity $n$ 
we will have the predicate \solname{$C$}{$f$} that
implements the function $f$. The arity of this predicate is $n+1$
where $n$ is the arity of the function (remember that member functions
in \SLAM\ are functions with an element of the class as the first
argument). 
Predicates \prename{$C$}{$f$} and \postname{$C$}{$f$} will implement
the \emph{checking} predicates of the pre and postconditions of the
function $f$.

\begin{figure}[htb]
\begin{tabbing}
$\translate (SP) =$
     \= $\bigcup_{rule~\in~SP} \transrule (rule, rule.class)~\cup$ \\
     \> $\bigcup_{\small{\miss} (f, C)} \transinh (f, C, \defines (f, C)) ~\cup$ \\
     \> $\bigcup_{C~\in~SP} \transread (C) ~\cup$ \\
     \> $\bigcup_{C~\in~SP} \transtrav (C) ~\cup$ \\
     \> $\transpred$ \\
\\
$\transrule (r, C) = \transpre (r,C) ~\cup~ \transpost (r,C) ~\cup~ \transsol (r,C)$\\
$\transpre (r, C) =  \prename{C}{f}(\transterm (C, r.args)) ~\texttt{:-}~ \transexpr (r.pre, \texttt{Pre}, C)\texttt{, Pre == true.}$ \\
$\transpost (r, C) =$ \=$\postname{C}{f}(\transterm (C, r.args), \texttt{Result}) ~\texttt{:-}~ \transexpr (r.post, \texttt{Post}, C)\texttt{,}$\\
\>\hspace{0.5cm}{\tt Post == true.}\\
$\transsol (r, C) = \solname{C}{f}(\transterm (C, r.args), \texttt{Result}) ~\texttt{:-}~ \transexpr (r.sol, \texttt{Result}, C)\texttt{.}$\\
$\transinh (f, C, C') = \bigcup_{a \in \mbox{C}} \texttt{\solname{C}{f}(}\transterm (C', a)\texttt{, R) :-} \solname{C}{f}\texttt{(}\transterm (C', a)\texttt{, R).}$\\
\\
$\transexpr ($\SLAMexpr{c ($e_1$, \ldots, $e_n$)}$, \texttt{V}, C) =$ \=
          $\transexpr (e_1, \texttt{X}_1, C)\texttt{,} \ldots\texttt{,} 
          \transexpr\ (e_n, \texttt{X}_n, C)\texttt{,}$\\
          \>$\texttt{V is } \transterm (C, c) \texttt{(}\texttt{X}_1, \ldots, \texttt{X}_n\texttt{).}$\\
$\transexpr ($\SLAMexpr{f ($e_1, \ldots, e_n$)}$, \texttt{V}, C) =$ \=
          $\transexpr (e_1, \texttt{X}_1, C)\texttt{,} \ldots\texttt{,} 
          \transexpr\ (e_n, \texttt{X}_n, C)\texttt{,}$\\
          \>$\solname{C}{f}\texttt{(}\texttt{X}_1, \ldots, \texttt{X}_n, \texttt{V).}$ \\
$\transexpr ($\SLAMexpr{obj.f ($e_1, \ldots, e_n$)}$, \texttt{V}, C) = \transexpr ($\SLAMexpr{f (obj, $e_1, \ldots, e_n$)}$, \texttt{V}, C)$ \\
$\transexpr ($\SLAMexpr{C':f ($e_1, \ldots, e_n$)}$, \texttt{V}, C) = \transexpr ($\SLAMexpr{f ($e_1, \ldots, e_n$)}$, \texttt{V}, C')$ \\
$\transexpr ($\SLAMexpr{\SLAMResult}$, \texttt{V}, C) = \texttt{V is Result.}$\\
$\transexpr ($\SLAMexpr{$e_1$ log\_op $e_2$}$, \texttt{V}, C) =$ \=
          $\transexpr (e_1, \texttt{X}_1, C)\texttt{,} 
          \transexpr (e_2, \texttt{X}_2, C)\texttt{,}$\\ 
          \>$\texttt{log\_op (X}_1\texttt{, X}_2\texttt{, V).}$\\
$\transexpr ($\SLAMexpr{\SLAMQFExpr{\aquan}{$x$}{$coll$}{{\it filt}}{$e$}}$, \texttt{V}, C) = $ \quant{\aquan}\texttt{(}$coll$, {\it filt}, $e$, \texttt{V)}
\end{tabbing}
\caption{Formalization of the translation.} \label{trans}
\end{figure}

The complete translation function \translate\ (see figure \ref{trans})
accepts a \SLAM\ program as argument and returns a Prolog program. The
definition uses several auxiliary functions.

\begin{myitemize}
  
\item \transrule\ accepts a rule and the class where the rule appears
  and returns three Prolog clauses: for \solname{$C$}{$f$},
  \prename{$C$}{$f$} and \postname{$C$}{$f$}. In the definition a rule
  is represented by a record with six fields: \textit{class} (the
  class where it is defined), \textit{fname} (function name),
  \textit{args} (arguments), \textit{sol} (the solution proposed by
  the rule), \textit{pre} (checkable part of the precondition), and
  \textit{post} (checkable part of the postcondition).
  
\item \transsol, \transpre, and \transpost\ are responsible for
  producing clauses for the function implementation and for checking
  the pre and postconditions.
  
\item \transexpr\ produces Prolog atoms in order to translate a \SLAM\ 
  expression. \transexpr ($e$, \texttt{X}, $C$) ensures that the Prolog
  variable \texttt{X} stores the computation of the expression $e$ in
  the class $C$.
  
\item \transinh\ is designed to produce the clause associated to an
  inherited function $f$ definition in class $C$ checked by \miss.
  \defines\ returns the ancestor class where a function is declared (see
  below for further details).
  
\item \transterm\ translates a term in a class into the corresponding
  Prolog representation (following the name policy described above).
  
\end{myitemize}

In the compilation scheme presented in figure \ref{trans} 
$c$ represents an attribute
constructor while $f$ represents a function. $a \in C$ \emph{produces}
all possible skeletons of terms in class $C$. {\it log\_op} is any
\SLAM\ logical operator computed in Prolog by the corresponding
\texttt{log\_op} predicate\footnote{Notice that they cannot be 
implemented by the similar Prolog operations (conjunction $\leftrightarrow$ {\tt ,}, etc.)
because \SLAM\ boolean functions compute \SLAMexpr{false} as a valid
result. We leave their definitions as a
  trivial exercise.}.  \transpred\ contains a Prolog implementation of
\SLAM\ predefined operations. Anyway the scheme deserves a deep
discussion on some aspects:

It is important to note that the same predicate name is used to
compute the same function in all the classes coming from the
ancestor class. Some few words are needed to justify this decision and
to show that it really handles the overloading produced by the use of
inheritance. As Prolog has no types, a predicate can be applied to any
number of functors on the top of arguments. Remember that attribute 
constructors in different classes have different names. If $f$ is a
member function of $C$ and the functions is applied to an expression
$e$ the generated Prolog atoms are:

\begin{center}
  $\transexpr(x,\texttt{Y},C)$\texttt{, \solname{$C$}{$f$}(Y,R)}
\end{center}

Notice that the clauses of \solname{$C$}{$f$} includes all the
definitions of $f$ in the complete class hierarchy, so it does not
matter the exact subclass $A$ of $C$ which $x$ belongs to. As
\texttt{Y} will not contain free variables and the  attribute constructors in all
the classes have different names, only the adequate original rule is
applicable. A similar approach is used for \postname{$C$}{$f$} and
\prename{$C$}{$f$}.

Patterns are translated into Prolog patterns, so parameter
passing is handle by unification. This use of unification is quite
efficient in Prolog because our programs only use the \emph{read mode}
of the underlying WAM machine. Function composition is handled by
flattening the functional structure and using auxiliary variables to
connect the predicates.

There are additional technical details that must be solved. One
appears when 
a function in class $A$ is fully inherited from a class
$C$, i.e. there are no rules for $f$ in $A$. As the attribute constructors in
both classes may be different, we need to add a clause to
transforms the arguments to have the constructors in $C$ and then call
recursively the same predicate. Suppose that the attribute constructor in $C$ is
\SLAMexpr{c} and the attribute constructor in $A$ is \SLAMexpr{c'} with possibly more
attributes. The generated clause is:

\begin{center}
 \solname{$C$}{$f$}\texttt{(c'($\gargs{x}$,$\gargs{y}$), R) :- } 
                     \solname{$C$}{$f$}\texttt{(c($\gargs{x}$), R).}
\end{center}

  The second one is how to compile a function call with explicit
  indication of the class. This expression could involve a type
  casting from an argument to an ancestor class. For each class $C$ we
  could generate a predicate \casting{$C$} that transforms all the
  elements of any successor class $A$ into $C$.  A call of the form
  \SLAMexpr{obj.$C$:f ($\gargs{x}$)} is translated into the atoms

  \begin{center}
    \casting{$C$}\texttt{(obj,Obj),}
    \solname{$C$}{$f$}\texttt{(Obj,$\gargs{x}$,R)}
  \end{center}

Notice that the dispatch table used for the compilation of inheritance is
simulated here by the use of different functors names (keys into
the table) as well as the unique predicate with all the possible
implementations of the function (the table itself).  One can argue
that in case of an intensive use of inheritance this huge predicate
can be a source of inefficiency. Remember that the Prolog program is
just used for debugging, so efficiency is not a crucial point. On the
other hand, this possible source of inefficiency is partially solved
by the indexing included in WAM-based Prolog implementations.

Translation of traversals and quantified expressions follow an
interesting approach. The \transtrav\ translation scheme
is applied for any class \SLAMexpr{C} implementing the
\SLAMexpr{TravCollection} and consists in the generation of
 a predicate \inname{C}. The
first argument of the predicate is the traversable object, the second
argument \emph{produces} by backtracking all the values in the
collection. A simple traversal is translated in this way:

\begin{small}
\begin{tt}
\begin{tabbing}
\inname{C}(O,X) :- first(O,X).\\
\inname{C}(O,X) :- next(O,O2), inside(O2), \inname{C}(O2,X).
\end{tabbing}
\end{tt}
\end{small}
\noindent
while a traversal definition using a sequence of traversals is
translated into several clauses, one for each traversal in the sequence:

\begin{small}
\begin{tt}
\begin{tabbing}
\inname{C}(<<shape>>,X) :- \inname{C$^1$}(<<tr$_1$>>,X).\\
\ldots\\
\inname{C}(<<shape>>,X) :- \inname{C$^n$}(<<tr$_n$>>,X).
\end{tabbing}
\end{tt}
\end{small}

A trivial translation of a quantifier definition is got by using the
following predicate quantifier. \ciao\ allows a special syntax {\tt f
  ($\mathsf{\gargs{A}}$) := {\it Expr} :- {\it condition}} to compute
the function {\tt f} as the expression {\tt\it Expr} provided that
{\tt\it condition} holds. The \ciao\ higher order contains the usual
higher order functions, like \verb.~.{\tt foldr($f, s, [x_1, \ldots, x_n]$)}
that computes the value of $f (x_1, f (x_2, \cdots f (x_n, s)\cdots)$.
There is also a special syntax for lambda expressions (called
predicate abstractions) {\tt \labs ($\gargs{\mathsf{X}}$, R) :- Expr},
where {\tt R} stores the computed expression {\tt Expr} applied to the
arguments $\gargs{\mathsf{X}}$.  We show the implementation
of some quantifiers as an example.

\begin{small}
\begin{tt}
\begin{tabbing}
quantifier(Binop,Seed,Collection,Filter,Expression) := \verb.~.foldr(Binop,Seed,L) :-\\
\hspace{1cm}\=    findall(Y,(in(Collection,X),Filter(X),Expression(X,Y)),L). \\
\quant{all}(Collection,Filter,Expression, R) :- \\
\>  R = quantifier(log\_and,true,Collection,Filter,Expression).\\
\quant{sum}(Collection,Filter,Expression, R) :- \\
\>  R = quantifier(+,0,Collection,Filter,Expression).\\
\quant{map}(Collection,Filter,Expression, R) :- \\
\>  R = quantifier(\labs (X, L, R) :- R = [X|L],[],Collection,Filter,Expression).
\end{tabbing}
\end{tt}
\end{small}

In some cases the filter predicate can generate values in a very
efficient way and a more efficient translation could be
\texttt{(Filter(X), in(Collection,X))} in the condition of the ``findall´´
construct. In principle, we have not information for choosing between
both alternatives, unless we use some techniques borrowed
from program analysis.

It is worth mentioning that, by this translation, 
it will be clear that the \SLAM\ selection quantifier provides
the full power of Prolog search, what enhances the expressive power
of \SLAM.

Notice that the checkable part of function pre and postconditions are
translated also into special predicates.  The only tricky question is
the use of the same argument patterns of the function instead of the
variables to ensure the adequate parameter passing even for the
conditions.

The last part of the translation that must be explained is the goal of
the $\transread\ (C)$ clauses. For each class $C$ a predicate with
name \readname{C} is defined in order to read an element of class $C$
from a class. The format of the data in the file is exactly the same
that is stored by the counterpart operation (\texttt{serialize}) of
the C++ class for $C$.  Due to the lack of space, we omit the
definition although it is an easy exercise.

Let us show the Prolog code for our example. Again we have omitted some
parts, and included
some simplifications to make it more readable, like the implementation
of sequences as lists or the inclusion of intermediate predicates to
compute the quantified expressions.

\begin{small}
\begin{tt}
\begin{tabbing}
amount(tran(S,D,A),A). \\
\solname{Bank}{MakeBank}(N, S, bank(N,S)). \\
sumSource(Ctr,N,V) :-
\quant{sum}(CTr,(\labs:(tran(S,D),R) :- S == N -> R = true; R = false),  amount, V). \\
sumDest(Ctr,N,V) :-
\quant{sum}(CTr,(\labs:(tran(S,D),R) :- D == N -> R = true; R = false),  amount, V). \\
\solname{CTransaction}{FinalAmount}(CTr, Banks, Result) :- \\
\hspace{1cm} \=\quant{map}(Banks, true, \labs \=(N, R) :-  \\
\>\>sumSource(CTr, N, A1), sumDest(CTr, N, A2), A is A1 - A2, \\
\>\>R = \solname{Bank}{MakeBank}(N,A)).\\
\prename{CTransaction}{FinalAmount}(CTr, Banks) :-
    Length(Banks, L), L > 0. \\
\postname{CTransaction}{FinalAmount}(CTr, Banks, Result, Post) :- \\
\>  Length(Banks, L1),
    Length(Result, L2),
    L1 == L2, \\
\>  all(1..L1, true, (\labs (I, R) :- \=nth (Result, I, bank (N, A)), nth (Banks, I, N), \\
\>\>                                     sumSource (CTr, N, A1), sumDest (CTr, N, A2), \\
\>\>                                     A == A1 - A2), true). \\
\inname{Seq}([X|\_], X). \\
\inname{Seq}([\_|L], X) :- \inname{Seq}(L, X).\\
\inname{Range}(N .. M,N). \\
\inname{Range}(N..M,R):- N <= M, NN = N+1, \inname{Range}(NN..M, R).
\end{tabbing}
\end{tt}
\end{small}

It is obvious that the transformation can be improved avoiding several
sources of inefficiency: generation of useless intermediate variables,
predicate calls that can be folded, construction of the result in the
head of the clause, etc. However, it is worth mentioning that the
\ciao\ system includes several optimizations based on partial
evaluation and global analyses in the compilation process that can
solve part of the inefficiencies. The system can eliminate part of the
conditions to be checked (\cite{german-phd,ciaopp-iclp99-tut}).

\section{Execution of assertions}
\label{check}

We have developed a C++ library called \texttt{"check.h"} including
the macros needed to execute the assertions. Basically, the library
contains three operations: \texttt{pre-check (}\textit{function name,
  arguments}\texttt{)}, to check preconditions,
\texttt{post-check(}\textit{function name, arguments,
  result}\texttt{)}, to check postconditions, and
\texttt{post-return-check(}\textit{function name, arguments,
  result}\texttt{)}, to check a postcondition and return a value.

The interface between C/C++ and Prolog is very primitive. A string
containing the goal is used to call the Prolog program. The arguments
(and the result, in the case of the postconditions) 
are stored in a file \texttt{\textit{file}} via the
\texttt{serialize} operation. A process is spawned to run the Prolog
program, which reads from the file the arguments/result and calls the
corresponding predicate. The pseudo-code for \texttt{post-return-check
  (}\textit{fname, $\gargs{args}$, result}\texttt{)} is the following:

\begin{small}
\begin{tt}
\begin{tabbing}
r = result; \\
fd = fopen (\textit{file}); \\
$arg_1$.serialize (fd); \\
\vdots, \\
$arg_n$.serialize (fd); \\
r.serialize (fd); \\
fclose (fd); \\
fork (prolog (\postname{C}{fname}); \\
assert (\textit{return code of process}); \\
return r;
\end{tabbing}
\end{tt}
\end{small}

\noindent
The other cases are analogous. We only need to add the Prolog program
two clauses for each function $f$ of the form:

\begin{small}
\begin{tt}
\begin{tabbing}
\prename{C}{f} :- \readname{$C_1$}(A$_1$), \ldots, \readname{$C_n$}(A$_n$),\prename{C}{f}(A$_1$, \ldots, A$_n$).\\
\postname{C}{f} :- \= \readname{$C_1$}(A$_1$), \ldots, \readname{$C_n$}(A$_n$),\readname{$C_r$}(Result),\\
\>\postname{C}{f}(A$_1$, \ldots, A$_n$, Result).
\end{tabbing}
\end{tt}
\end{small}

The use of the file for exchange information between the C++ program
and the Prolog program can be optimized using a pipe to link the processes.
As a pipe is handled as a file, there is no significant difference w.r.t the
previous scheme.

There is an alternative possibility. First of all, we modify
\texttt{serialize} to construct a Prolog representation of data as a
string.  Then a Prolog goal of the form:

\begin{center}
\begin{small}
\begin{tt}
\begin{tabbing}
assert (arguments ($arg_1$.serialize (), \ldots $arg_n$.serialize (), \textit{result}.serialize ()), \\
arguments (A$_1$, \ldots, A$_n$, Result), \postname{C}{f} (A$_1$, \ldots, A$_n$, Result).
\end{tabbing}
\end{tt}
\end{small}
\end{center}

\noindent
is used to call the Prolog engine.

The Prolog \emph{assert} is needed to ensure that the arguments are
compiled.  The main problem with this approach is that when the data
are large, the string representation of the goal is longer than the
accepted size by the Prolog compiler.

Additionally, the debugging execution produces a report on the
pre/postconditions checked, indicating those conditions that where
marked as a subformula (\SLAMAndCheck) or alternative condition
(\SLAMEitherCheck) of the full pre/postcondition.

\section{Future work and Conclusion}
\label{future}

We have presented the validation facilities of the \SLAM\ system. From
a high level specification the system generates code to execute the
system including assertions to check pre and postconditions of the
program operations. These high level logical formulae are compiled into
Prolog and the resulting program is responsible for checking them. The
compiled \SLAM\ program is linked with the assertion Prolog program to
achieve the debugging facilities that are specially useful when the
programmer decides to modify the automatically generated code.

The overall efficiency is acceptable. The execution of the program in
the debug mode needs between 3-4 times the execution of the program
itself.

The \SLAM\ project is at the early stages of its development. The
specification language is not fully defined yet and the system
(environment, compilers, libraries, etc.) is under development.  We
plan to have a prototype with the debugging facilities by the date of 
the workshop.

The \SLAM\ project and its debugging facilities seem to be a very
useful tool to develop high quality programs, i.e. error free with
respect to the specification, clean, easy to read and manipulate to
achieve modifications either in the specification or in the generated
code, fully documented and including high level declarative debugging
facilities to allow optimizations in a reliable way.

The debugging facilities can be used for other purposes. It is easy to
modify the system to allow assertions along the code, for instance to
check loop invariants.

For the behaviour of the system is case of failure of the debugging assertions
we rely on the underlying debug facilities (the effect of \texttt{assert} in C++
in our case). Of course it is possible to modify this behaviour to use
failures to locate more precisely the bug and the wrong function in the
vein of algorithmic debugging \cite{Shapiro82}. More precisely, when a function call
violates its postcondition, function invocations inside the code
are checked. Those that have an incomplete postcondition (i.e.
annotated as \SLAMAndCheck or \SLAMEitherCheck) are on suspicion
of wrong behaviour.

Another relatively easy extension is to use the \SLAM\ debugging facilities
for program validation by testing automatically sequences of data designed
to cover all the cases.

\ToDo{
\section{Acknowledgements}

We are grateful to \ldots}


\end{document}